\relax
\documentclass[letterpaper]{article} 
\usepackage{aaai22}  
\usepackage{times}  
\usepackage{helvet}  
\usepackage{courier}  
\usepackage[hyphens]{url}  
\usepackage{graphicx} 
\usepackage{natbib}  
\usepackage{caption} 
\DeclareCaptionStyle{ruled}{labelfont=normalfont,labelsep=colon,strut=off} 
\usepackage{algorithm}
\usepackage{algorithmic}
\usepackage{amsfonts}
\usepackage{amstext}
\usepackage{amsmath}
\usepackage{booktabs}
\def\ouralgo{{\text{NES}}}
\usepackage{newfloat}
\usepackage{listings}
\floatstyle{ruled}
\newfloat{listing}{tb}{lst}{}
\floatname{listing}{Listing}
\title{Learning Large-scale Network Embedding from Representative Subgraph}

\author {
    Junsheng Kong\textsuperscript{\rm 1}\thanks{Work was done during internship at Tencent Quantum Lab.},
    Weizhao Li\textsuperscript{\rm 1},
    Ben Liao\textsuperscript{\rm 2}\equalcontrib,
    Jiezhong Qiu\textsuperscript{\rm 3},
    Chang-Yu (Kim) Hsieh\textsuperscript{\rm 2},
    Yi Cai\textsuperscript{\rm 1}\equalcontrib, 
    Jinhui Zhu\textsuperscript{\rm 1},
    Shengyu Zhang\textsuperscript{\rm 2}
}

\affiliations {
    \textsuperscript{\rm 1} School of Software Engineering, South China University of Technology\\
    \textsuperscript{\rm 2} Tencent Quantum Lab\\
    \textsuperscript{\rm 3} Tsinghua University\\
    sescut\_kongjunsheng@mail.scut.edu.cn, 
    bliao@tencent.com,
    ycai@scut.edu.cn
}

\begin{document}

\maketitle

\begin{abstract}
We study the problem of large-scale network embedding, which aims to learn low-dimensional latent representations for network mining applications. 
Recent research in the field of network embedding has led to significant progress such as DeepWalk, LINE, NetMF, NetSMF. 
However, the huge size of many real-world networks makes it computationally 
expensive to learn network embedding from the entire network.
In this work, we present a novel network embedding method called ``NES", which learns network embedding from a small representative subgraph. NES leverages theories from graph sampling to efficiently construct representative subgraph with smaller size which can be used to make inferences about the full network, enabling significantly improved efficiency in embedding learning. Then, NES computes the network embedding from this representative subgraph, efficiently.
Compared with well-known methods, extensive experiments on networks of various scales and types demonstrate that NES achieves comparable performance and significant efficiency superiority. 
\end{abstract}

\section{Introduction}
Networks are ubiquitous in the real world such as social network~\citep{aggarwal2011introduction, myers2014information}, citation networks~\citep{sen2008collective}, biological networks~\citep{zitnik2019evolution}, chemical networks~\citep{martins2012bayesian}. Mining information from the real network plays a crucial role in many emerging applications.
In recent years, the emergence of network embedding technology provides a revolutionary paradigm for modeling graphs and networks.
Network embedding methods aim to automatically learn the low-dimensional latent representation of each node in networks. A large number of research have shown that network embeddings can capture structural attributes of the network, and they are effective for downstream network applications such as vertex classification, image classification, link prediction and recommendation systems~\cite{DBLP:conf/kdd/DongCS17, DBLP:conf/www/ZhangWZ19, DBLP:conf/kdd/DongZTCW15}.

Briefly, current research on network embedding can be divided into three categories: (1) skip-gram based method, these methods are inspired by Word2vec~\citep{sgns}, such as LINE\citep{tang2015line}, DeepWalk~\citep{perozzi2014deepwalk}, Node2vec~\citep{grover2016node2vec}, Metapath2vec\cite{DBLP:conf/kdd/DongCS17} and VERSE\cite{DBLP:conf/www/TsitsulinMKM18}; (2) deep neural networks based methods, such as \cite{DBLP:conf/iclr/KipfW17, DBLP:conf/kdd/YingHCEHL18}; (3) matrix factorization based methods such as GraRep~\citep{cao2015grarep}, NetMF~\citep{qiu2018netmf} and NetSMF~\citep{qiu2019netsmf}. Among them, NetMF unifies the network embedding methods based on skip-gram into the matrix factorization framework.

Although these methods have made good progress in learning network embedding, they still have limitations. Especially in large-scale network embedding learning task, the computational cost of these methods increases sharply with the increase of the network scale. Methods based on skip-gram and deep neural networks are costly to train. For example, with default parameters set, DeepWalk takes several months to embed an academic collaborative network of 67 million vertices and 895 million edges. Node2Vec performs a high-order random walk and takes more time to learn embedding than DeepWalk. Although matrix factorization based method (NetSMF) is more efficient than the above two methods, it still needs to perform highly expensive information matrix construction and factorization operations on the large-scale matrix.

\begin{figure*}[t]
\centering
\includegraphics[width=0.75\textwidth]{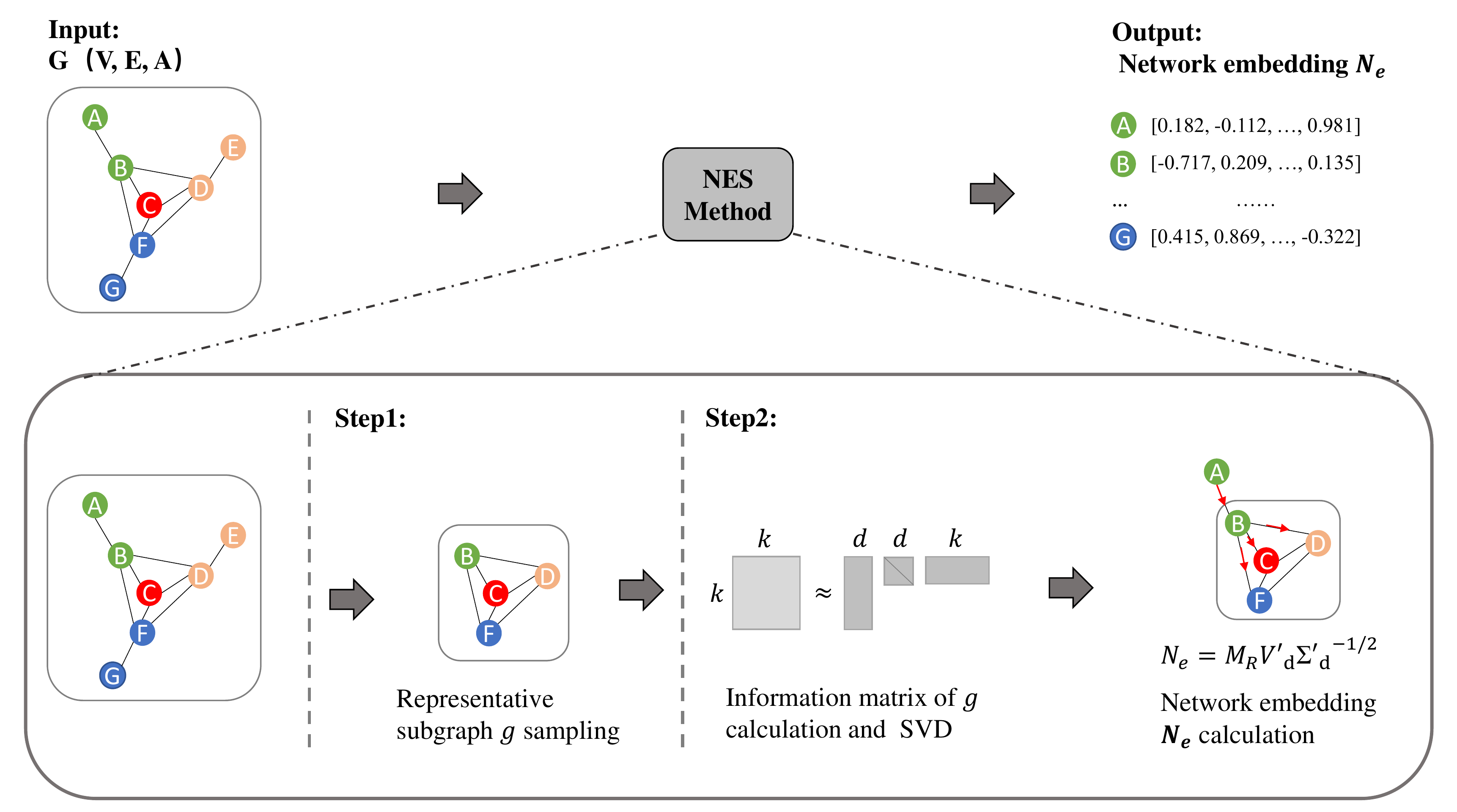} 
\caption{$\ouralgo$ method.}
\label{Fig_model}
\end{figure*}

To tackle this issue, we get inspiration from the matrix sampling algorithm~\cite{DBLP:journals/jacm/FriezeKV04}. 
We find that the representative subgraph in the original large network are sufficient to build a representation with good performance. By sampling the typical nodes of the original large-scale network, we obtain a small-scale and representative subgraph. Different from matrix factorization based methods which need to construct and factorize the whole original large network, we only need to carry out information matrix construction and factorization in the small representative subgraph, thus reducing the computing resources and ensuring the performance of network embeddings. The algorithm scheme is shown in Fig.~\ref{Fig_model}. First, it samples the small representative subgraph from the original large network through degree-based node sampling. Secondly, it learns network embedding of the whole network from this small representative subgraph (including construction and factorization of the small information matrix, embedding calculation).

The key contributions of this paper are as follows:
\begin{itemize}
    \item We develop our $\ouralgo$ method. $\ouralgo$ makes it available to embed the whole network with low computational complexity and memory requirement. In our method $\ouralgo$, the complexity of information matrix construction and factorization is relevant only to the number of nodes in a small representative subgraph. Since the representation of nodes in our method only depends on the relationship between nodes and representative subgraphs, our method has the scalability to super-large networks.
    \item Through extensive experiments on three real-world network datasets of different scales and topics, we present that the proposed $\ouralgo$ achieves orders of magnitude speedup over various state-of-the-art methods, and maintains equivalent or better accuracy at the same time.
\end{itemize}

\section{PRELIMINARIES}
We define the input network as $G=(V, E, A)$, where $V=\{v_1,v_2,...v_n\}$ is a collections of $n$ nodes and $E=\{e_{i,j}\}_{i,j=1}^n$ is the edge set.
Specifically, $e_{i,j}=1$ if there exists an edge between the node $v_i$ and $v_j$ otherwise $e_{i,j}=0$. We use an adjacent matrix $A \in \mathcal{R}^{n \times n}$ to represent connects of a full network. The goal of network embedding method is to learn a function $f : V \to  \mathcal{R}^{d}$ that maps each vertex $v_i$ to a $d$-dimensional vector while capturing structural properties of the graph. 

This paper focuses on the matrix factorization based method for network embedding. Previous work~\cite{qiu2018netmf} has shown that well-known network embedding methods (DeepWalk, LINE, PTE, node2vec) are in essence implicitly factorizing a matrix with a closed form, and embeddings generated by the matrix factorization based methods can get comparable performance or even better performance. It reveals that DeepWalk\citep{perozzi2014deepwalk}
essentially factorizes a matrix derived from the random walk process. Inspired by previous research~\cite{DBLP:conf/nips/LevyG14}, it also proves that when the length of random walks goes to infinity, DeepWalk implicitly and asymptotically factorizes the information matrix as follows:
\begin{equation}
M = \log_+\big (\frac{vol(G)}{b\cdot T}\sum_{r=1}^{T} (D^{-1}A)^{r}D^{-1}\big),
\label{information_matrix}
\end{equation}
where $vol(G)=\sum_i \sum_j A_{ij}$ denotes the volume of the graph, $\log_+$ is the entry-wise truncated logarithm defined as $\log_+(x) =
max(\log(x), 0)$, and
$D = diag(d_1, d_2, d_3,...,d_n)$ denotes the degree matrix with $d_i = \sum_{j=1}^{n}A_{ij}$.
Finally, we can factorize information matrix $M$ by using Singular Value Decomposition (SVD) and construct network embedding $N_e$ by using its top-$d$ singular values/vectors as follows:
\begin{equation}
    U_d, \Sigma_d, V_d ~=~tSVD(M),
\end{equation}
\begin{equation}
    N_e = U_d \Sigma_d^{1/2}. 
\end{equation}

However, directly constructing and factorizing this dense matrix is extremely time-consuming. To reduce the computational cost, NetSMF sparsifies the aforementioned dense information matrix, enabling significantly improved efficiency in embedding learning. Although NetSMF is more efficient than the NetMF, it still needs to factorize a large $n \times n$ sparse matrix, where $n$ is the number of nodes. This makes it highly expensive to directly factorize and calculate for large-scale network embedding. In this paper, we propose a novel network embedding method called ”NES”, which learns network embedding from the representative subgraph. NES leverages theories from graph sampling to efficiently construct representative subgraph with smaller size which can be used to make inferences about the full network, enabling significantly improved efficiency in embedding learning. 
We list the notations and their descriptions in Table.~\ref{tab:notation}.

\begin{table}[htbp]
\caption{Notation.}
\centering
\setlength{\tabcolsep}{1.8mm}{
\begin{tabular}{c|l}
\toprule
\textbf{Notation}                & \textbf{Description}                \\ \midrule
$G$            & input network           \\
$V$            & vertex set of G with $|V|=n$          \\
$E$            & edge set of G with $|E|=m$           \\
$A$            & adjacency matrix of $G$           \\
$D$            & degree matrix of $G$           \\
$vol(G)$            & volume of $G$           \\
$d$            & dimension of embedding           \\
$b$            & number of negative samples           \\
$T$            & context window size           \\
$g$            & representative subgraph of input network $G$           \\
$A_g$            & adjacency matrix of $g$           \\
$\tilde A_g$            & $\sum_{i=1}^T A_g^{i}$           \\
$R$            & related matrix           \\
$M$            & information matrix of $G$           \\
$M_g$            & information matrix of $g$           \\
$I_k$            & identity matrix with size of $k$           \\
$N_e$            & node embedding matrix of $G$           \\

\bottomrule
\end{tabular}
}
\label{tab:notation}
\end{table}

\section{Method}
In this section, we present $\ouralgo$ method —— an efficient and effective
method for large-scale network embedding learning problem. We develop the $\ouralgo$ method to construct and factorize a small representative information matrix that approximates the original information matrix. The main idea of our method $\ouralgo$ is to find a representative subgraph with smaller size which can be used to make inferences about the full network. Both the construction and the factorization of the small information matrix require a low cost. With this design, we are able to demonstrate running-time supremacy for solving a large-scale network embedding problem and maintain performance. In this section, we will first introduce the connection between network embeddings and the representative subgraph. Then, we will describe the process of our method $\ouralgo$ in detail.


\begin{figure}
    \centering
    \includegraphics[scale=0.45]{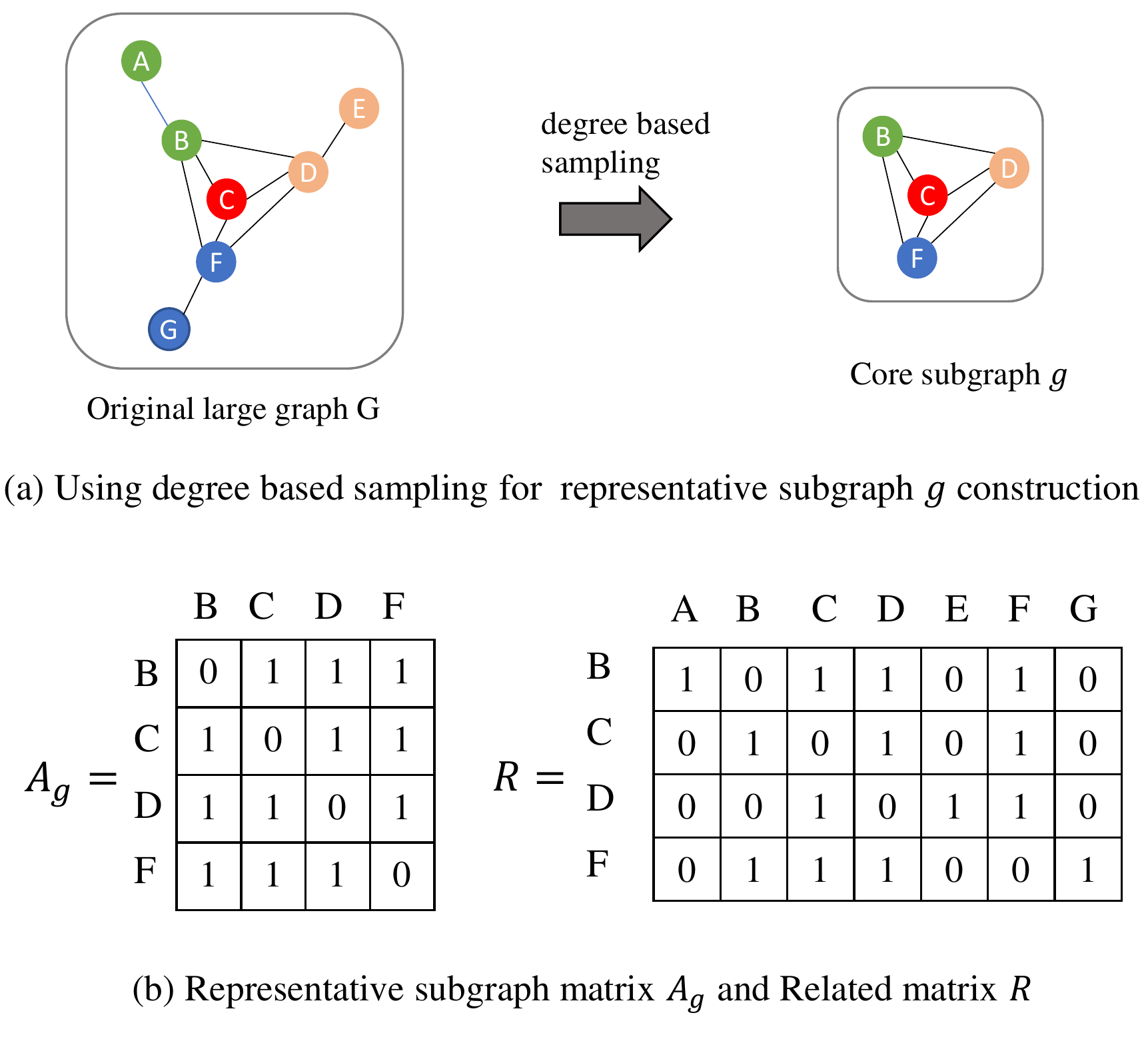}
    \caption{(a) shows the process of representative subgraph sampling. (b) presents the representative subgraph matrix $A_g$ and related matrix $R$.}
    \label{fig:subgraph}
\end{figure} 

\subsection{Connection between network embeddings and representative subgraph}
As shown in preliminaries section, we know that network embedding matrix
$N_e$ can be computed by the form $N_e=U_d\Sigma_d^{1/2}$,
where $U_d$ and $\Sigma_d$ are top-$d$ left singular vectors and top-$d$ singular values of information matrix $M$ of $G$. Since $M \approx U_d \Sigma_d V_d^{\top}$, network embeding $N_e$ also can be obtain by the form $$N_e = U_d\Sigma_d^{1/2} = U_d \Sigma _d V_d^{\top} V_d \Sigma^{-1/2} \approx M V_d\Sigma_d^{-1/2}.$$ 
For each node embedding $e_i = M_i V_d\Sigma_d^{-1/2}$. This form reveals the relation between a target node embedding with the contextual nodes in whole network. 
Previous work~\cite{DBLP:journals/jacm/FriezeKV04} shows that with high probability: 
\begin{equation*}
    M^\top M\approx S^\top S
\end{equation*}
where matrix $S$ picks $k$ rows of matrix $M$ independently at random. 
It proves that the SVD of $M$ from the spectral decomposition (SD) of $M^{\top}M$ can be approximate from the SD of small and typical $S^{\top}S$. Sampling for each row is equivalent to sampling a node of the network. Previous work\cite{ebbes2016sampling} has also shown that some nodes with high degrees reflect the presence of some influential nodes. Inspired by these, we use degree-based node sampling to get small and representative subgraph $g$. To get approximate $V_d^{'}$ and $\Sigma_d^{'}$, we decompose the information matrix $M_g$ of representative subgraph $g$. The information matrix $M_g$ with $T$ steps walk can be computed like Eq.\ref{information_matrix}. 
We reveal a simple relation between node embeddings with the representative subgraph $g$ as follows:
\begin{equation}
N_e =  M_R V_d^{'}\Sigma_d^{'-1/2}
\label{eq:ne}
\end{equation}
where $M_R$ denotes the information matrix between all nodes in network with nodes in representative subgraph $g$.

\subsection{Step 1: Representative Subgraph sampling}
We first focus on the goal of sampling a representative subgraph. The term ``representative subgraph” refers to producing a small sample of the original network, whose characteristics represent as accurately as possible the entire network\cite{DBLP:conf/kdd/LeskovecF06}. As shown in the preliminaries section, matrix factorization based methods need to construct and factorize the large $n \times n$ information matrix $M$. To scale down this information matrix $M$, we apply node sampling to construct a smaller representative subgraph with a size of $k$. 

Previous research shows that node sampling based methods can sample a representative subgraph such as degree based sampling\cite{DBLP:conf/kdd/LeskovecF06}, PageRank weight, ICLA-NS\cite{DBLP:journals/eaai/GhavipourM17}. For a large-scale network, it is essential to use a particularly efficient sampling method. Previous work shows that the networks have some nodes with high degrees reflecting the presence of some influential nodes\cite{ebbes2016sampling}. Based on these, we choose the degree based sampling to construct a representative subgraph with size of $k$ as shown in Fig.~\ref{fig:subgraph}. In our method, we first compute the degree of all nodes, then we select nodes with the $k$ highest degrees to construct the representative subgraph $g$. In our experiment, the result indicates this sampling method is efficient and effective. Through this step, we get the representative subgraph $g$ and the related matrix $R$ which includes the relationship between all nodes with representative nodes $c$. We summarize this process in Algorithm~\ref{alg-sub}.


\begin{algorithm}[tb]
\caption{subgraph sampling method}
\textbf{Input}: input network $G = (V,E,A)$ which we need to learn network embedding $N_e$, $|V|=n$; \\
\textbf{Parameter}: number of samples $k$;  \\
\textbf{Output}: adjacency matrix $A_g$ of representative subgraph $g$, related matrix $R$
\begin{algorithmic}[1] 
\STATE Select nodes with the $k$ highest degrees to get representative nodes set $c$.
\STATE Construct the related matrix $R \in \mathbb{R}^{k \times n}$: $R_{i,*} = A_{c_i,*}$.
\STATE Construct the representative matrix $A_g \in \mathbb{R}^{k \times k}$ : ${A_g}_{*,i} = R_{*,c_i}$.
\STATE Normalize the related matrix $R$ and representative subgraph matrix $A_g$.

\STATE \textbf{return} related matrix $R$, representative matrix $A_g$.
\end{algorithmic}
\label{alg-sub}
\end{algorithm}

\subsection{Step 2: Network embedding from Representative Subgraph}
After constructing the representative subgraph $g$, we  calculate network embeddings from subgraph $g$.

\textbf{Step 2.1: Construct an information Matrix sparsifier from subgraph.}
Both in NetMF and NetSMF methods, it needs to construct and factorize a large $n \times n$ information matrix, where the large $n$ is the number of nodes. Different from these, NES only needs to construct and factorize the smaller $k \times k$ information matrix $M_g$ of representative subgraph $g$. Given a subgraph matrix $A_g$, we first compute matrix power from $A_g^1$ to $A_g^T$ and then get $\tilde A_g$. However, this process will make $\tilde A_g$ to be a dense matrix with $O(k \times k)$ number of non-zeros. To reduce the construction cost, we use Random-Walk Molynomial Sparsification~\cite{DBLP:conf/colt/ChengCLPT15, DBLP:journals/corr/ChengCLPT15} to get a sparse matrix $\tilde A_g$ like NetSMF~\cite{qiu2019netsmf}. Then, we construct the information matrix $M_g$ of subgraph as follows:

$$M_g~=~\log_+(\frac{vol(G)}{T\cdot b} \tilde A_g).$$

\textbf{Step 2.2: Truncated singular value decomposition.}
The next step is to perform truncated singular value decomposition (tSVD) on the constructed information matrix $M_g$. In this work, we apply randomized tSVD (rtSVD) to factorize the information matrix $M_g$. Randomized tSVD has been shown to be efficient \cite{DBLP:journals/siamrev/HalkoMT11}. 
We factorize the information matrix $M_g$ of subgraph as follows:

$$U_d^{'}, \Sigma_d^{'}, V_d^{'} ~=~rtSVD(M_g).$$
where $\Sigma_d^{'}$ is the diagonal matrix formed from the top-$d$ singular values, and $U_d^{'}$ and $V_d^{'}$ are $k \times d$ orthonormal matrices corresponding to the selected singular values. Through this step, we get $M_g \approx U_d^{'} \Sigma_d^{'} V_d^{'\top}$.
\begin{algorithm}[tb]
\caption{$\ouralgo$  method}
\textbf{Input}: input network $G \in \mathbb{R}^{n \times n}$; \\
\textbf{Parameter}: number of samples $k$, and embedding dimension $d$; \\
\textbf{Output}: Network embedding $N_e$
\begin{algorithmic}[1] 
\STATE Construct the subgraph matrix $g \in \mathbb{R}^{k \times k}$ and related matrix $R \in \mathbb{R}^{k \times n}$ according to Algorithm~\ref{alg-sub};
\STATE Use Random-Walk Molynomial Sparsification to get a sparse matrix $\tilde A_g$ to approximate the sum of matrix power from $A_g^1$ to $A_g^T$.

\STATE Compute information matrix $M_g$ from sub-graph matrix $\tilde A_g$:
$M_g~=~\log_+(\frac{vol(G)}{T\cdot b} \tilde A_g).$
\STATE Factorize the information matrix $M_g$ by randomized tSVD: $U_d^{'}, \Sigma_d^{'}, V_d^{'} ~=~rtSVD(M_g).$

\STATE Compute the information matrix for $R$:\\
$M_R~=~\log_+(\frac{vol(G)}{T\cdot  b} R^{\top} \cdot (\tilde A_g +I_k)).$

\STATE Compute the network embedding $N_e \in \mathbb{R}^{n \times d}$:\\ $N_e~=M_RV_d^{'}\Sigma_d^{'-1/2}.$

\STATE \textbf{return} $N_e$
\end{algorithmic}
\label{alg-main}
\end{algorithm}

\textbf{Step 2.3: Network embedding computation.} 
From the analysis of connection between network embeddings and representative subgraph, we know that network embedding matrix
$N_e$ can be computed from subgraph $g$ by the form $$N_e =  M_R V_d^{'}\Sigma_d^{'-1/2},$$
where $V_d^{'}$ and $\Sigma_d^{'}$ are top-$d$ right singular vectors and top-$d$ singular values of information matrix $M_g$ of subgraph $g$, $M_R$ denotes the information matrix between all nodes in network with nodes in representative subgraph. 

To get approximate $M_R$, we can use related matrix $R$ and $\tilde A_g$ to compute $M_R$ as follows:
$$M_R~=~\log_+(\frac{vol(G)}{T\cdot b} R^{\top} \cdot (\tilde A_g +I_k)).$$



\textbf{Complexity Analysis.} 
We get Algorithm~\ref{alg-main} by putting the above procedures together. 
As for line 1, it requires $\mathcal O(n)$ time to construct the degree of all nodes. As for line 2, line 3 and line 4, time complexity for Random-Walk Molynomial Sparsification and information matrix construction are only related to the sampling size $k$.
Both in NetMF and NetSMF methods, it needs to construct and factorize a large $n \times n$ information matrix, where the large $n$ is the number of nodes. 
As for line 5, $\mathcal O(nk^{2})$ time is spent in matrix multiplication.
As for line 6, $\mathcal O(kdn)$ time is spent in matrix multiplication.
In total, the computation complexity of $\ouralgo$ is linearly dependent on the magnitude of size of nodes $n$.


\begin{figure*}
    \centering
    \includegraphics[scale=0.6]{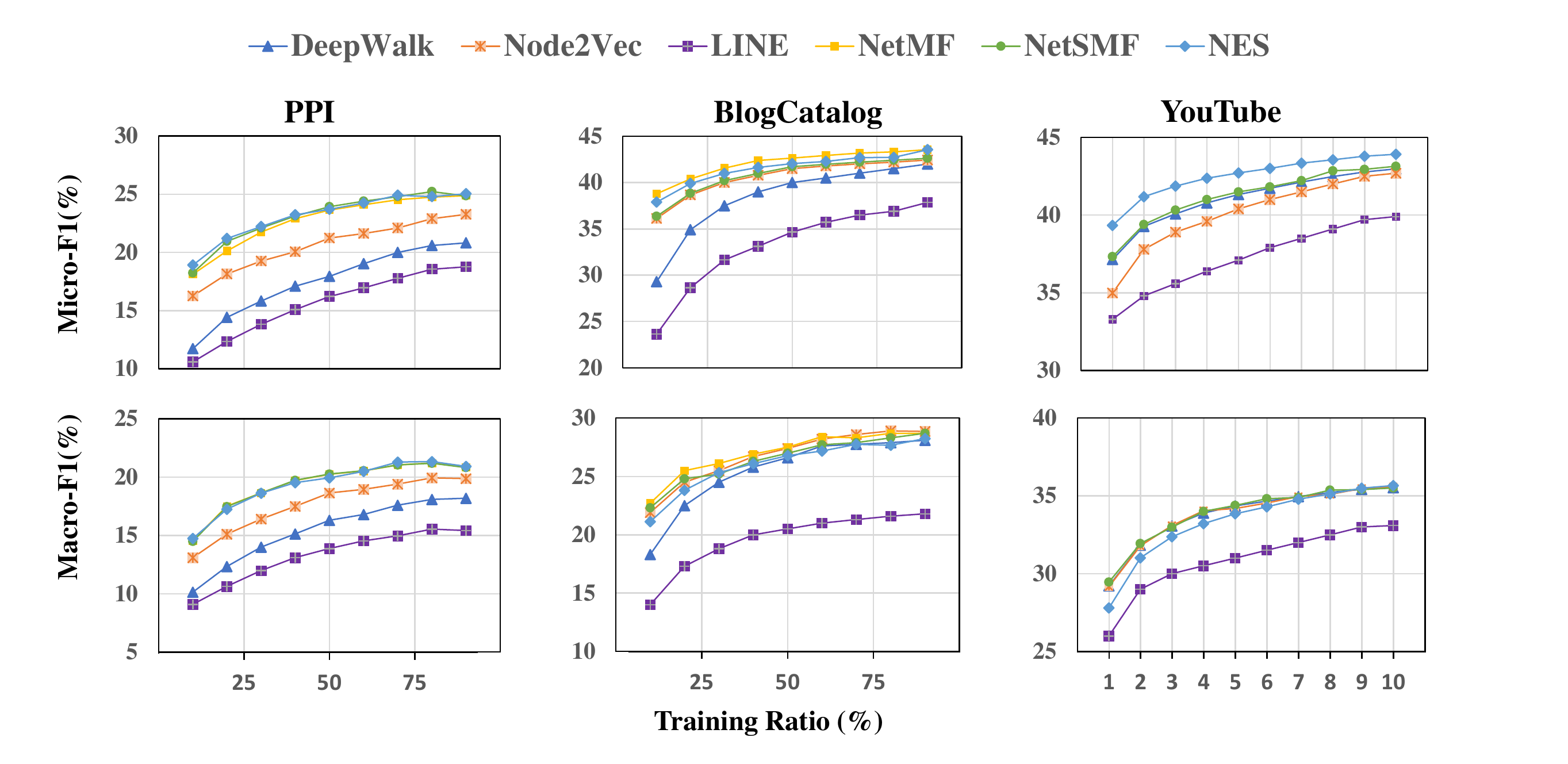}
    \caption{Prediction performance on varying the ratio of training data. The $x$-axis represents the ratio of labeled data (\%), and the $y$-axis in the top and bottom rows denote the Micro-F1 and Macro-F1 scores respectively.}
    \label{fig:total_result}
\end{figure*}

\section{Experiment}
In this section, we evaluate the proposed NES method on the multi-label vertex classification task, which has been commonly used to evaluate previous network embedding techniques\cite{perozzi2014deepwalk,tang2015line,qiu2019netsmf}. 

\subsection{Datasets}

\begin{table}[ht]
\centering
\begin{tabular}{cccc} 
\toprule
dataset & \#nodes   & \#edges   & \#labels  \\ 
\hline
PPI     & 3,890     & 76,584    & 50        \\
BlogCatalog    & 10,312    & 333,983   & 39        \\
Youtube & 1,138,499 & 2,990,443 & 47        \\
\bottomrule
\end{tabular}
\caption{Statistics of Datasets.}
\label{tab:datasets}
\end{table}

We employ three datasets for the vertex classification task, which have been widely used in network embedding literature, including BlogCatalog, PPI and YouTube. The statistics of these datasets are listed in Table~\ref{tab:datasets}.

\begin{itemize}

\item \textbf{Protein-Protein Interactions (PPI)}~\citep{dataset:ppi} is a subgraph of the PPI network for Homo Sapiens. The vertex labels are obtained from the hallmark gene sets and represent biological states.

\item \textbf{BlogCatalog}~\citep{dataset:blog,dataset:blog1} is a network of social relationships of online bloggers. The vertex labels represent the interests of the bloggers.

\item \textbf{YouTube}~\citep{dataset:youtube} is a video-sharing website that allows users to upload, view, rate, share, add to their favorites, report, comment on videos. The users are labeled by the video genres they liked.
\end{itemize}

\begin{figure*}
    \centering
    \includegraphics[scale=0.47]{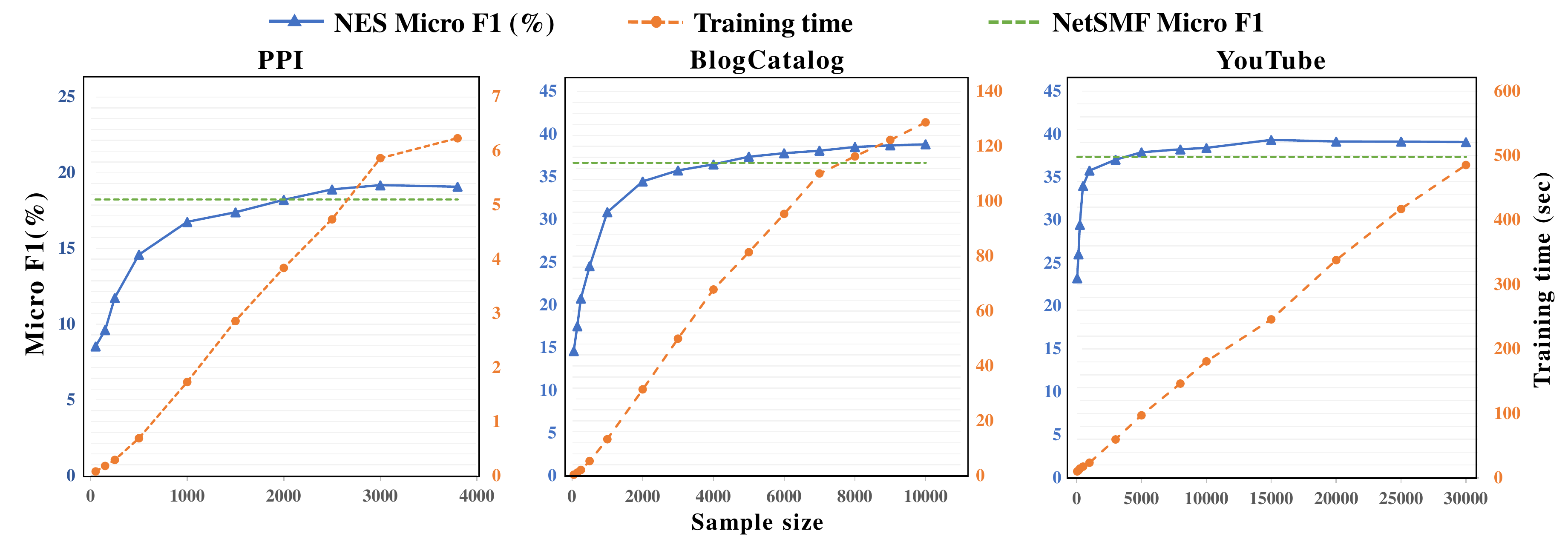}
    \caption{The micro F1 scores and the corresponding training times of {\ouralgo} method under different sample sizes. The size of the original network are about 3,890 (on PPI), 10,312 (on BlogCatalog) and 1,138,499 (on YouTube) respectively.}
    \label{fig:sampling}
\end{figure*}

\subsection{Baselines}
To verify the performance of $\ouralgo$, we compare it with several state-of-the-art methods. Methods and their parameters are briefly introduced below. We use the same parameters as in the original paper. Across all datasets, we set the embedding dimension $d$ to be 128. The context window size $T$ is set to be 10. All experiments are carried out on a cloud server with Intel(R) Xeon(R) Gold 6231C CPU.

\begin{itemize}
    \item \textbf{DeepWalk}~\citep{perozzi2014deepwalk} learns node embedding with local information obtained from truncated random walks. We set the walk length $t=40$ and walks per vertex $\gamma=80$.
    \item \textbf{Node2Vec}~\citep{grover2016node2vec} designs a biased random walk strategy to explores diverse neighborhoods efficiently. It is optimized with grid search over its return and in-out parameters $(p,q) \in \{0.25, 0.5, 1, 2, 4\}$.  The walk length and number of walks per vertex remain the same as DeepWalk.
    \item \textbf{LINE}~\cite{tang2015line} is trained by optimizing an objective function of edge reconstruction. We set negative sampling edge size to be 5, and edge sample size to be $10^{10}$. 
    \item \textbf{NetMF}~\citep{qiu2018netmf} explicitly factorize the closed-form
matrices that DeepWalk and LINE aim to implicitly approximate
and factorize. The number of eigen pairs is set as 256, and number of negative samples is set as 1.
    \item \textbf{NetSMF}~\citep{qiu2019netsmf} is a variant of NetMF and leverages theories from spectral sparsification to speed up computation.
    \item \textbf{NES} is our proposed method. We set the sample size $k=2,500$ for PPI, $k=6,000$ for BlogCatalog and $k=15,000$ for YouTube.
\end{itemize}


\subsection{Predcition Setting}
We measure the quality of embeddings following the same experimental procedure in DeepWalk~\citep{perozzi2014deepwalk}. 
We randomly selected a portion of the labeled nodes to train a classifier. The rest of the nodes are used for testing.
For PPI and BlogCatalog, the training ratio is varied from 10\% to 90\%. For YouTube, the training ratio is varied from 1\% to 10\%. The classifier is set as one-vs-rest logistic regression model implemented by LIBLINEAR~\citep{fan2008liblinear}. We report the average Micro-F1 and Macro-F1 scores for all methods.

\begin{table}[htbp]
\centering
\caption{Efficiency comparison based on running time (including filesystem IO and training computation). ``×'' indicates that the corresponding algorithm is unable to handle the computation due to excessive memory.}
\begin{tabular}{c|ccc} 
\toprule
         & PPI     & BlogCatalog & YouTube    \\ 
\hline
DeepWalk & 4 mins  & 12 mins     & 1 day      \\
Node2Vec & 4 mins  & 56 mins     & 4 days     \\
LINE     & 41 mins & 40 mins     & 46 mins    \\
NetMF    & 16 secs & 2 mins      & ×          \\
NetSMF   & 10 secs & 13 mins     & 4.1 hours  \\
NES      & 4.7 secs  & 1.5 mins    & 4.1 mins   \\
\bottomrule
\end{tabular}
\label{tab:time_compare}
\end{table}

\subsection{Experimental Results}

To demonstrate the speed advantage of $\ouralgo$,  we compare the running time of different baselines as shown in Table~\ref{tab:time_compare}. 
Our method is significantly faster than other baselines. 
Remarkably, $\ouralgo$ can embed the YouTube (large-scale network) with about 4 minutes while the fastest baseline LINE is at least 11 times slower and NetSMF is about 60 times slower. The skip-gram baed baselines (DeepWalk and Node2Vec) require more than one day to train such large network. NetMF performs the dense approximation on the whole network, making it infeasible for YouTube dataset.
Similar speedups can be consistently observed from BlogCatalog (moderate-size network) and PPI (small network).
The efficiency of LINE drops dramatically in small-scale network since the redundant edge reconstruction operations. 

We also summarize the prediction performance of all methods on three datasets in Fig.~\ref{fig:total_result}. 
Comparing with baselines, results obtained with $\ouralgo$ are fairly close to the ones obtained with NetMF and sometimes even better, despite the fact we only use a fraction of the original network. 
In PPI, our $\ouralgo$ is relatively indistinguishable from NetMF and NetSMF.
In BlogCatalog, our $\ouralgo$ has slightly worse performance than NetMF but better performance than NetSMF in terms of Micro-F1.  
In YouTube, we can see that $\ouralgo$ has a significant advantage over other methods in terms of Micro-F1. As the ratio of training data increases, NES can also obtain the same scores as other methods in terms of Macro-F1.
With the node sampling strategy, $\ouralgo$ can capture the representative information about the whole network.



The analysis presented above confirms that $\ouralgo$~method can substantially minimize the training time while maintaining benchmark performance.

\subsection{Discussion and Analysis}

\begin{table}[htbp]
\centering
\caption{The training time decomposition of the different part of {$\ouralgo$} method on three networks. The training time of {$\ouralgo$} on PPI, BlogCatalog and YouTube are 4.7 seconds, 95.4 seconds and 245.9 seconds respectively.}
\vskip 0.1in
\begin{tabular}{cccc}
\toprule
                            & \multicolumn{1}{l}{\begin{tabular}[c]{@{}l@{}}Subgraph\\ sampling\end{tabular}} & \multicolumn{1}{l}{\begin{tabular}[c]{@{}l@{}}Network embedding\\ form representative subgraph\end{tabular}} \\ \hline
PPI                         & 1.46\%                                                                          & 98.54\%                                                                                         \\
BlogCatalog                 & 0.33\%                                                                          & 99.67\%                                                                                         \\
\multicolumn{1}{l}{YouTube} & 3.43\%                                                                          & 96.57\%                                                                                         \\ 

\bottomrule
\end{tabular}
\label{tab:time_dis}
\end{table}

\textbf{Time analysis.}
Recall two main steps of the \ouralgo~method: representative subgraph sampling, network embedding from representative subgraph (including construction of  information  matrix, computation of singular values and left singular vectors, and embedding computation). The breakdown of computational time is displayed in Table.\ref{tab:time_dis}. 
Note that the Subgraph sampling step takes up only no more than 4\% of the whole time, 
which shows the efficiency of Subgraph sampling step.

\textbf{Sample size.}
In our method, the representative subgraph is obtained by node sampling from the original large network. To investigate how the sample size affects the performance of $\ouralgo$, we carry out experiments with different sample sizes on different datasets as shown in Fig.\ref{fig:sampling}. 
To determine the impact of different sample sizes, we have fixed the training data ratio as 10\% for PPI and BlogCatalog, and 1\% for YouTube during prediction. 
As the sample size increases, it is obvious that the performance of $\ouralgo$ gradually increases and surpasses that of NetSMF. 
The convergence is fairly fast, especially when the network scale is large. We can see that to exceed the performance of NetSMF, the sample size of $\ouralgo$ only needs to be 2,500 (on PPI), 6,000 (on BlogCatalog) and 15,000 (on YouTube), which are about 64.27\%, 58.18\% and 1.32\% of the number of nodes in the original graph, respectively.
We also observe diminishing gains for too large sample size. We conjecture that representative subgraph with proper sample size can provide main network information. 
As the scale of network becomes larger, the percentage of sampling nodes decreases. In terms of training time, it increases linearly as the number of samples. This indicates that the training time can be effectively reduced when sampling a small subgraph of the original graph.

\begin{figure}
    \centering
    \includegraphics[scale=0.37]{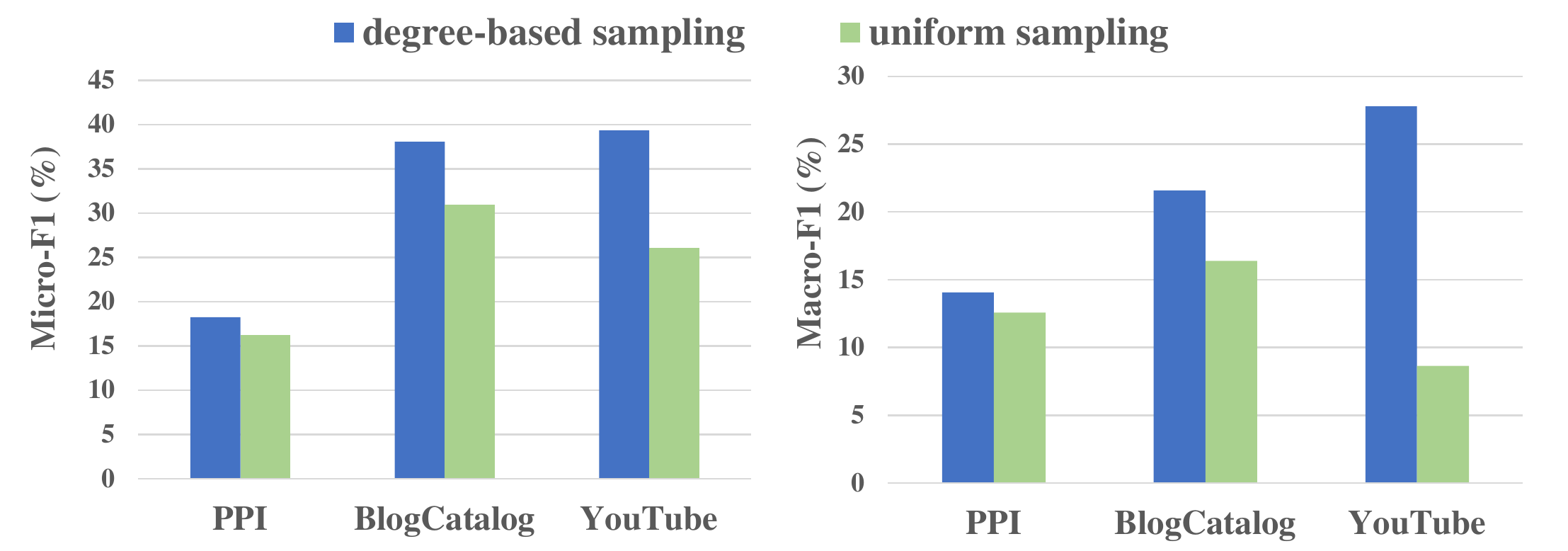}
    \caption{Comparison between degree-based sampling and uniform sampling.}
    \label{fig:uniform_sampling}
\end{figure}

\textbf{Degree-based sampling vs uniform sampling.}
In $\ouralgo$ method, we use degree-based sampling to capture the typical information in the original large-scale network. To verify the effectiveness of the degree-based sampling, we replace the degree-based sampling with uniform sampling in the sub-graph construction step. Uniform sampling means that each node in the network will be sampled with equal probability. The compared results are shown in Fig.~\ref{fig:uniform_sampling}.
It shows that our degree-based sampling has great advantages over uniform sampling. In terms of micro F1 score, degree-based sampling has 11\% $\sim$ 34\% increase compared with uniform sampling. In term of macro F1 score, degree-based sampling has 11\% $\sim$ 69\% increase compared with uniform sampling. As the network scale enlarges, our degree-based sampling can gain more advantages over the uniform sampling.

\section{Related Work}
In this section, we review the related work of network embedding and graph sampling.

\textbf{Network embedding.} 
Network embedding has been extensively studied over the past years~\citep{DBLP:journals/debu/HamiltonYL17} and it is widely used in a lot of downstream network applications, such as recommendation systems~\citep{ying2018graph}. In general, network embedding learns latent low-dimensional feature representations for the nodes or edges in a network. Traditional dimension reduction methods can naturally be applied to obtain network embedding through Singular Value Decomposition (SVD)~\citep{ou2016asymmetric,wang2017community}. Inspired by the skip-gram~\citep{sgns} algorithm in word embedding, DeepWalk~\citep{perozzi2014deepwalk} and Node2Vec~\citep{grover2016node2vec} generate truncated random walks over a network. The random node sequences are regarded as sentences and fed into a language model to get the embedding. NetMF~\citep{qiu2018netmf} unifies a collection of skip-gram based network embedding methods into a matrix factorization framework.

Most of the existing network embedding methods cannot be directly applied to large-scale networks. 
In recent years, many research efforts have been dedicated to improving network embedding methods to accommodate large-scale network computation.
LINE~\citep{tang2015line} is proposed to gain scalability on large networks, which preserves the first and second order proximities. 
NetSMF~\citep{qiu2019netsmf} leverages theories from spectral sparsification to efficiently sparsify the dense matrix in the NetMF framework.
MILE~\citep{MILE} coarsens the graph in several iterations, gradually reduces the size of the graph and computes the network embedding of the smallest graph.

\textbf{Graph sampling.} 
The goal of graph sampling is to efficiently estimate the graph properties by picking a subset of nodes/edges from the original graph.
The graph sampling technology arises in many applications such as graph visualization~\citep{graphVisual}, survey hidden population~\citep{hiddenpopulation}. 
In our $\ouralgo$, we apply node sampling strategy.
Nodes can be sampled uniformly without re-placement~\citep{relate:RN}, proportional to the degree centrality of nodes~\citep{relate:RDN} or according to the pre-calculated PageRank score of the vertices~\citep{relate:PRN}.



\section{Conclusion}
In this work, we focus on the problem of large-scale network embedding. 
Different from previous work, we present the algorithm of Large-scale Network embedding from Representative Subgraph ($\ouralgo$). $\ouralgo$ method only needs to construct and factorize information matrix of the representative subgraph with small size. With this design, $\ouralgo$ is able to learn network embedding efficiently from large-scale network with million nodes. Extensive experiments on networks of various scales and types demonstrate that $\ouralgo$ achieves both effectiveness and significant efficiency superiority when compared to the well-known baselines.

\bibliography{main}

\bigskip

\end{document}